# Partially coherent conical refraction

**V.Yu. Mylnikov,[1,*] V.V. Dudelev,[2] E.U. Rafailov[3] and G.S. Sokolovskii [1,4]**

[1]*Ioffe Institute, 26 Polytechnicheskaya str., St. Petersburg, 194021, Russia*
[2]*Peter the Great St. Petersburg Polytechnic University, St. Petersburg, 195251, Russia*
[3]*Optoelectronics and Biomedical Photonics Group, AIPT, Aston University, Aston Triangle, Birmingham, B4 7ET, UK*
[4]*St. Petersburg Electrotechnical University (LETI), St. Petersburg, 197022, Russia*
*\*Vm@mail.ioffe.ru*

**Abstract:** In this paper, we extend the paraxial CR model to the case of the partially coherent light using the unified optical coherence theory. We demonstrate the decomposition of CR correlation functions into well-known CR coherent modes for a Gaussian Schell-model source. Assuming randomness of the electrical field phase of the input beam, we reformulated and significantly simplified the rigorous CR theory. This approach allows us to consider the propagation of light through a CR crystal in exactly the same way as in the classical case of coherent radiation. Having this in hand, we derive analytically the CR intensity both in the focal plane and in the far field, which allows us to explain and rigorously justify earlier experimental findings and predict new phenomena. The last include the counterintuitive effect of narrowing of the CR ring width, disappearance of the dark Poggendorff's ring in the Lloyd's plane, and shift of Raman spots for the low-coherent CR light. We also demonstrate a universal power-law dependence of CR cones coherence degree on the input correlation length and diffraction-free propagation of the low-coherent CR light in the far field.

## 1. Introduction

In 1832, Hamilton introduced concept of the internal conical refraction (CR) [1], that can be observed after the propagation of the collimated light beam along one of the optical axis of the biaxial crystal. After the exit crystal facet the radiation emerges as a light ring [2] with radius $R_0$, defined by the CR crystal length and a product of its refractive indexes (so-called 'conicity', see Fig. 1(a)). A detailed study of the CR intensity distribution after the exit facet of a biaxial crystal reveals that the CR ring has a fine structure and consists of two bright concentric rings [3]. The outer ring is of greater intensity with a dark (Poggendorff) ring separating it from the inner dim one. Further spatial evolution of the CR beam (Fig. 1 (b)) leads to the formation of a bright axial spike, known as the Raman spot [4,5], which are found at each end of the axial CR distribution.

In general, there are three main characteristics in CR. First of all, CR has nonconventional beam evolution pattern [6], which is used for all-optical image processing [7], high-resolution microscopy [8–10], ultra-efficient CR lasing [11–14], and, moreover, for optical manipulation and actuation of physical objects and particle trapping / tweezing by means of CR ring distribution [15–18], Raman spots [16,18] and CR bottle beam structure [19–21]. Secondly, this is a spatially nonuniform polarization structure [22], which can be beneficial for optical polarization converters [23], multiplexed systems [24], formation of mono- or polychromatic structured light [25–27], polarimetric measurements [28–30] and optical sensors [31]. And thirdly, CR wavefront possess the nontrivial helical properties associated with fractional orbital angular momentum (OAM) [32]. This CR feature provides important practical applications such as generation of singular beams and annihilation of optical vortices [33–36], generation of sum-frequency [37] and high harmonics [38–40] (see also the recent review [41]). It is noteworthy that, despite the two-century history of this optical phenomenon, CR still attracts a lot of attention and continues to impress the optical community with many new findings [42–48].

The utilization of nontrivial optical beams for CR is now one of the most novel and intriguing areas of research. Recent publications report on the CR with a top-hat [49], Dark /Antidark [50], Bessel [51] and Bessel-Gauss [52] beams, elegant Laguerre–Gaussian modes [53], multimode radiation [54], and vector beams with spatially varying state of polarization [55]. In this regard, the study of CR with partially coherent light becomes an important and promising direction for further research. From a practical point of view, this is very beneficial due to the properties of low-coherent light sources such as laser diodes and LEDs, which are simple, compact and very cost-effective. Furthermore, spatial coherence provides an additional opportunity to control the properties of light, aside from the unique polarization, propagation and phase characteristics of the

CR. In this way, one can generate so-called structured light which is currently of great interest for theoretical and experimental studies [56]. Examples of coherence control-enabled structured light include partially coherent vector beams [57], which combine unconventional features of polarization and coherence, and coherence vortices [58], which are found when coherence is taken into account for OAM beams. As these beams combine unique features of vector/vortex and low-coherent beams, their generation by means of partially coherent CR can have an immediate impact for many application of coherence control-enabled structured light including optical coherence encryption [59], free space optical communications [60] and beam shaping [61]. The first steps in the study of partially coherent CR were taken in a recent paper [62], which demonstrates dramatic dependence of the characteristic features of CR intensity pattern on the spatial coherence of the input light beam. However, a rigorous theory of this phenomenon hasn't yet been developed.

In this paper, we extend the paraxial CR model to the case of the partially coherent light using the unified optical coherence theory. We demonstrate the decomposition of CR correlation functions into well-known CR coherent modes for a Gaussian Schell-model source. Assuming randomness of the electrical field phase of the input beam, we reformulated and significantly simplified the rigorous CR theory. This approach allows us to consider the propagation of light through a CR crystal in exactly the same way as in the classical case of coherent radiation. Having this in hand, we derive analytically the CR intensity both in the focal plane and in the far field, which allows us to explain and rigorously justify earlier experimental findings and predict new phenomena. The last include the counterintuitive effect of narrowing of the CR ring width, disappearance of the dark Poggendorff's ring in the Lloyd's plane, and shift of Raman spots for the low-coherent CR light. We also demonstrate a universal power-law dependence of CR cones coherence degree on the input correlation length and diffraction-free propagation of the low-coherent CR light in the far field.

## 2. Conical Refraction With Partially Coherent Radiation

Let us first consider the conical refraction (CR) formed by light, with the electric field vector in the form: $\mathbf{E}^{(0)}(\boldsymbol{\rho})=[E_L^{(0)}(\boldsymbol{\rho}), E_R^{(0)}(\boldsymbol{\rho})]=E^{(0)}(\boldsymbol{\rho})\mathbf{e}_{in}$, where $\boldsymbol{\rho}=\mathbf{r}/w_0$ is the transverse radius vector normalized to the beam waist $w_0$; $\mathbf{e}_{in}=[d_L, d_R]$ is the polarization vector, in which we choose left and right circular polarizations as basis functions; and $E^{(0)}(\boldsymbol{\rho})$ is a scalar function that determines the transverse distribution of the electric field. According to Belsky–Khapalyuk–Berry paraxial theory [63,64], after passing through the CR crystal, the electric field vector is transformed, as follows:

$$\mathbf{E}(\boldsymbol{\rho},\xi) = \begin{pmatrix} \mathcal{B}_0 & \mathcal{B}_{-1} \\ \mathcal{B}_1 & \mathcal{B}_0 \end{pmatrix} \begin{bmatrix} d_L \\ d_R \end{bmatrix}, \tag{1}$$

$$\mathcal{B}_\mu(\boldsymbol{\rho},\xi) = \int \frac{d\boldsymbol{\kappa}}{2\pi} \exp(i\boldsymbol{\kappa}\boldsymbol{\rho}) \tilde{G}_\mu(\boldsymbol{\kappa},\xi) \tilde{E}^{(0)}(\boldsymbol{\kappa}), \tag{2}$$

$$\tilde{G}_\mu(\boldsymbol{\kappa},\xi) = (-i)^\mu \cos(\kappa\rho_0 - \mu\tfrac{\pi}{2}) \exp\left[i\mu\theta_\kappa - i\kappa^2\xi/2\right], \tag{3}$$

where $\xi=z/k_0w_0^2$ is the longitudinal coordinate (1) normalized to the Rayleigh range of the incident beam; $k_0$ is a vacuum wavenumber; $\boldsymbol{\kappa}$ is the normalized transverse wave vector in cylindrical coordinates $(\kappa, \theta_\kappa)$; $\tilde{G}_\mu(\boldsymbol{\kappa},\xi)$ is a Green function describing both the transformation of the CR crystal and the spatial evolution along the $\xi$ axis; $\rho_0=R_0/w_0$ is a normalized radius of the CR ring beyond the crystal, which plays a central role in the determination of the resulting CR intensity distribution; and $\tilde{E}^{(0)}(\boldsymbol{\kappa})$ is the Fourier transform of the scalar function $E^{(0)}(\boldsymbol{\rho})$, defined as:

$$\tilde{E}^{(0)}(\boldsymbol{\kappa}) = \int \frac{d\boldsymbol{\rho}}{2\pi} \exp(-i\boldsymbol{\kappa}\boldsymbol{\rho}) E^{(0)}(\boldsymbol{\rho}). \tag{4}$$

Furthermore, we use the sign "~" to distinguish the Fourier transform of a function from the function itself.

From Equation (3), it follows that the CR field components $\mathcal{B}_\mu(\boldsymbol{\rho},\xi)$ have an extra orbital angular momentum (OAM), $\mu=0,\pm1$, since the Green function $\tilde{G}_\mu(\boldsymbol{\kappa},\xi)$ in (2) depends on the angle variable as $\exp(i\mu\theta_\kappa)$. When the input radiation is the simplest optical vortex, with the OAM equal to $m$ [35,36,53], the CR beam consists of several optical vortices, with an OAM equal to $m$, $m+1$, and $m-1$. Moreover, if the input beam is unpolarized, or circularly polarized, the CR radiation intensity can then be represented as the sum of the intensities with a different extra OAM:

$$I(\boldsymbol{\rho},\xi) = |\mathfrak{B}_0|^2 + \frac{1}{2}|\mathfrak{B}_1|^2 + \frac{1}{2}|\mathfrak{B}_{-1}|^2. \tag{5}$$

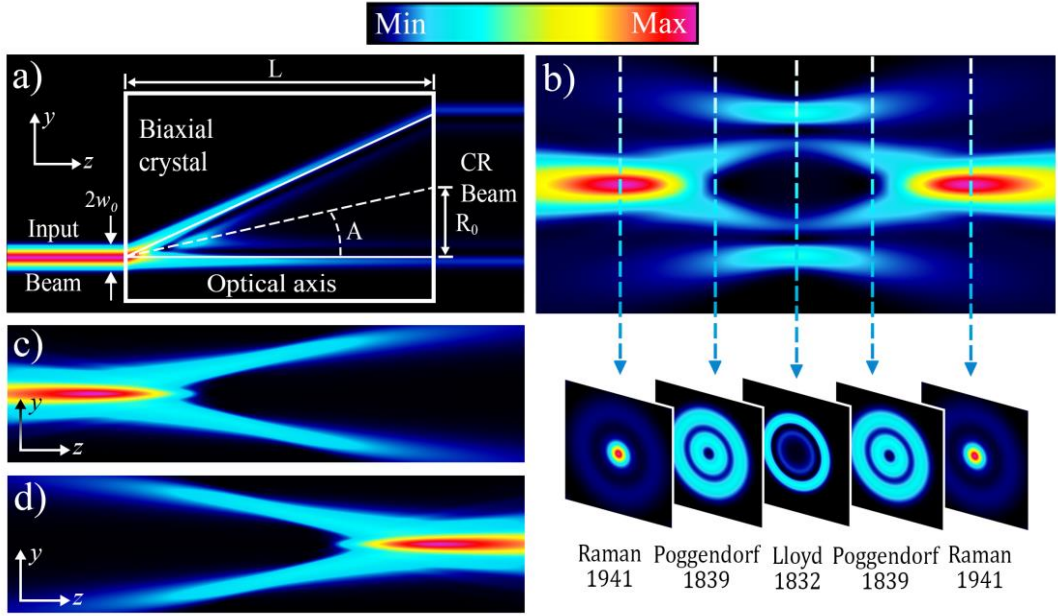

**Figure 1.** Spatial evolution of a collimated light beam through a conically refracting crystal of length, $L$: Inside the crystal, light transforms into a cone with a conicity, $A$; outside the crystal, the cone refracts into a cylinder of radius $R_0=AL$ (a). The axial evolution of the conical refraction (CR) beam and associated transverse intensity distributions are shown in (b); (c),(d) illustrate theoretical axial distributions of the individual CR cone $|C^{(+)}|^2$ and $|C^{(-)}|^2$, respectively.

Let us now introduce the coherence of the input beam in CR theory. Here, we will consider the input electric field vector $\mathbf{E}^{(0)}(\boldsymbol{\rho})$ as a random function. In this case, the set of correlation functions contains all of the information on the statistical properties of the field. Furthermore, we focus on the second-order correlation functions that can be used to construct a cross-spectral density matrix. The elements of this matrix are [65]:

$$\left\langle E_i^{(0)}(\boldsymbol{\rho}_1)^* E_j^{(0)}(\boldsymbol{\rho}_2) \right\rangle_{\text{All}} = W^{(0)}(\boldsymbol{\rho}_1,\boldsymbol{\rho}_2) J_{ij}, \tag{6}$$

where $\langle\ldots\rangle_{\text{All}}$ denotes the ensemble average over all possible realizations of the stochastic electric field vector, including averaging over both polarization and spatial degrees of freedom; $W^{(0)}(\boldsymbol{\rho}_1,\boldsymbol{\rho}_2)=\langle\mathcal{E}(\boldsymbol{\rho}_1)^*\mathcal{E}(\boldsymbol{\rho}_2)\rangle$ is an input cross-spectral density, which contains averaging only over the random scalar function $\mathcal{E}(\boldsymbol{\rho})$; and $J_{ij}=\langle d_i^* d_j\rangle_{\text{Polarization}}$ is a polarization matrix [65]. Equation (6) assumes that the light polarization is independent of the spatial degrees of freedom.

A cross-spectral density matrix, after the CR crystal is introduced in analogy to (6), is:

$$K_{ij}(\boldsymbol{\rho}_1,\boldsymbol{\rho}_2,\xi) = \left\langle E_i^*(\boldsymbol{\rho}_1,\xi) \cdot E_j(\boldsymbol{\rho}_2,\xi) \right\rangle_{\text{All}}, \tag{7}$$

where it is convenient to separate the polarization and spatial degrees of freedom, by analogy with Equation (6). Clearly, from Equation (1), any quadratic combination of electric field vector components of the form $E_i^* E_j$ can always be expressed in terms of different combinations of $B_\nu^* B_\mu$. As a result, for polarization-spatial separation in Equation (6), we need to introduce an additional set of correlations functions:

$$W_{\nu\mu}(\boldsymbol{\rho}_1,\boldsymbol{\rho}_2,\xi) = \left\langle \mathfrak{B}_\nu(\boldsymbol{\rho}_1,\xi)^* \mathfrak{B}_\mu(\boldsymbol{\rho}_2,\xi) \right\rangle, \tag{8}$$

similar to the input cross-spectral density $W^{(0)}(\boldsymbol{\rho}_1,\boldsymbol{\rho}_2)$ in Equation (6). We can call them orbital correlation functions since Equation (8) determines the correlation between the CR field components with an extra OAM equal to $v$ and $\mu$. The relationship between the introduced orbital functions (8) and the components of the cross-spectral density matrix (7) has the following form (Appendix 1):

$$\begin{pmatrix} K_{LL} \\ K_{RR} \\ K_{LR} \\ K_{RL} \end{pmatrix} = \begin{pmatrix} W_{00} & W_{-1-1} & W_{0-1} & W_{-10} \\ W_{11} & W_{00} & W_{10} & W_{01} \\ W_{01} & W_{-10} & W_{00} & W_{-11} \\ W_{10} & W_{0-1} & W_{1-1} & W_{00} \end{pmatrix} \begin{pmatrix} J_{LL} \\ J_{RR} \\ J_{LR} \\ J_{RL} \end{pmatrix}. \tag{9}$$

Thus, from formula (9), the CR intensity for an unpolarized input beam can be written as (Appendix 1):

$$I(\boldsymbol{\rho},\xi) = W_{00}(\boldsymbol{\rho},\boldsymbol{\rho},\xi) + \tfrac{1}{2}W_{11}(\boldsymbol{\rho},\boldsymbol{\rho},\xi) + \tfrac{1}{2}W_{-1-1}(\boldsymbol{\rho},\boldsymbol{\rho},\xi), \tag{10}$$

similar to formula (5), for coherent radiation.

It may also be useful to introduce partial spatial coherence into the dual-cone model of the CR [66,67]. Within its framework, the electric field vector behind the biaxial crystal can be represented as a sum of two CR cones, $\mathbf{E} = \mathbf{C}^{(+)} + \mathbf{C}^{(-)}$, where $\mathbf{C}^{(+)}$ is a positive or diverging cone (Figure 1c), while $\mathbf{C}^{(-)}$ is a negative or converging cone (Figure 1d), correspondingly.

$$\mathbf{C}^{(\pm)} = \begin{pmatrix} C_0^{(\pm)} & C_{-1}^{(\pm)} \\ C_1^{(\pm)} & C_0^{(\pm)} \end{pmatrix} \begin{bmatrix} d_L \\ d_R \end{bmatrix}, \tag{11}$$

$$C_\mu^{(\pm)}(\boldsymbol{\rho},\xi) = \int \frac{d\boldsymbol{\kappa}}{2\pi} \exp(i\boldsymbol{\kappa}\boldsymbol{\rho}) \tilde{G}_\mu^{(\pm)}(\boldsymbol{\kappa},\xi) \tilde{\mathcal{E}}(\boldsymbol{\kappa}), \tag{12}$$

$$\tilde{G}_\mu^{(\pm)}(\boldsymbol{\kappa},\xi) = \frac{1}{2}(\pm 1)^\mu \exp\left[\mp i\kappa\rho_0 + i\mu\theta_\kappa - i\kappa^2\xi/2\right]. \tag{13}$$

Orbital correlation functions for the dual-cone model are introduced, by analogy with (8):

$$V_{\nu\mu}^{(\sigma_1\sigma_2)}(\boldsymbol{\rho}_1,\boldsymbol{\rho}_2,\xi) = \left\langle C_\nu^{(\sigma_1)}(\boldsymbol{\rho}_1,\xi)^* C_\mu^{(\sigma_2)}(\boldsymbol{\rho}_2,\xi) \right\rangle, \tag{14}$$

with $\sigma_{1,2}=\pm$. As a result, the introduced functions (14) determine correlations between the CR cone component with index $\sigma_1$ and an extra OAM $v$, with respect to the component with index $\sigma_2$ and an extra OAM $\mu$. The relationship between correlation functions for Belsky–Khapalyuk–Berry theory (8) and the dual-cone model (14) is as follows:

$$W_{\nu\mu}(\boldsymbol{\rho}_1,\boldsymbol{\rho}_2,\xi) = \sum_{\sigma_1,\sigma_2=\pm} V_{\nu\mu}^{(\sigma_1\sigma_2)}(\boldsymbol{\rho}_1,\boldsymbol{\rho}_2,\xi), \tag{15}$$

where it is necessary to sum over all indices associated with the CR cones.

Let us now express the orbital correlation functions in terms of the input cross-spectral density. Substituting (2) into (8), for Belsky–Khapalyuk–Berry theory, we obtain the following relations:

$$W_{\nu\mu}(\boldsymbol{\rho}_1,\boldsymbol{\rho}_2,\xi) = \int \frac{d\boldsymbol{\kappa}_1 d\boldsymbol{\kappa}_2}{2\pi \, 2\pi} \exp[-i(\boldsymbol{\kappa}_1\boldsymbol{\rho}_1 - \boldsymbol{\kappa}_2\boldsymbol{\rho}_2)] \tilde{G}_\nu(\boldsymbol{\kappa}_1,\xi)^* \tilde{G}_\mu(\boldsymbol{\kappa}_2,\xi) \tilde{W}^{(0)}(\boldsymbol{\kappa}_1,\boldsymbol{\kappa}_2), \tag{16}$$

where $\tilde{W}^{(0)}(\boldsymbol{\kappa}_1,\boldsymbol{\kappa}_2)$ is a Fourier transform of the input cross-spectral density:

$$\tilde{W}^{(0)}(\boldsymbol{\kappa}_1,\boldsymbol{\kappa}_2) = \int \frac{d\boldsymbol{\rho}_1 d\boldsymbol{\rho}_2}{2\pi \, 2\pi} \exp[i(\boldsymbol{\kappa}_1\boldsymbol{\rho}_1 - \boldsymbol{\kappa}_2\boldsymbol{\rho}_2)] W^{(0)}(\boldsymbol{\rho}_1,\boldsymbol{\rho}_2). \tag{17}$$

A very similar formula can be derived by substituting (12) into (14) for the dual-cone model:

$$V_{\nu\mu}^{(\sigma_1\sigma_2)}(\boldsymbol{\rho}_1,\boldsymbol{\rho}_2,\xi) = \int \frac{d\boldsymbol{\kappa}_1 d\boldsymbol{\kappa}_2}{2\pi \, 2\pi} \exp[-i(\boldsymbol{\kappa}_1\boldsymbol{\rho}_1 - \boldsymbol{\kappa}_2\boldsymbol{\rho}_2)] \tilde{G}_\nu^{(\sigma_1)}(\boldsymbol{\kappa}_1,\xi)^* \tilde{G}_\mu^{(\sigma_2)}(\boldsymbol{\kappa}_2,\xi) \tilde{W}^{(0)}(\boldsymbol{\kappa}_1,\boldsymbol{\kappa}_2). \tag{18}$$

As a result, after numeric calculation of the integral (16) or (18), one can obtain full information on the second-order coherence effects [65] for the CR field. However, numeric evaluation of multidimensional integrals (16) or (18) requires large computational resources. Furthermore, we greatly simplify partially coherent CR theory by expressing the correlation functions in terms of the Belsky–Khapalyuk–Berry [63,64] and dual-cone model integrals [66,67].

## 3. The Coherent-Mode Representation of the Conical Refraction With Partially Coherent Radiation

In this section, we consider the Schell-model source [65]. Its cross-spectral density has the following form:

$$W^{(0)}(\mathbf{\rho}_1, \mathbf{\rho}_2) = \mathcal{E}^{(0)}(\mathbf{\rho}_1)^* \mathcal{E}^{(0)}(\mathbf{\rho}_2) g(\mathbf{\rho}_2 - \mathbf{\rho}_1), \tag{19}$$

where $\mathcal{E}^{(0)}(\mathbf{\rho})$ is the deterministic electric field amplitude in the focal plane, at a point with transverse coordinate $\mathbf{\rho}$; and $g(\mathbf{\rho}_2-\mathbf{\rho}_1)$ is a degree of spatial coherence. For a Gaussian Schell-model source, which we will consider, the field amplitude and the degree of coherence are of the form [65]:

$$\mathcal{E}^{(0)}(\mathbf{\rho}) = \exp\left[-\mathbf{\rho}^2/2\right], \quad g(\mathbf{\rho}_2 - \mathbf{\rho}_1) = \exp\left[-\Delta^2(\mathbf{\rho}_2 - \mathbf{\rho}_1)^2/2\right], \tag{20}$$

where the coherence parameter $\Delta = w_0/w_g$ is defined as the ratio of the focal beam waist $w_0$ to the characteristic scale of the degree of coherence or correlation length $w_g$. The coherence parameter serves as a measure of the "degree of global coherence" of the light source [65]. In the case of a small coherence parameter ($\Delta \ll 1$), the degree of coherence is approximately equal to unity, and remains almost unchanged, on the characteristic scale of the deterministic electric field amplitude. This means that the source may be said to be relatively coherent, in the global sense. Contrastingly, when a coherence parameter is large ($\Delta \gg 1$), the degree of coherence decreases on scales much smaller than the focal beam waist, $w_0$, i.e., the radiation source can be said to be relatively incoherent, in the global sense. It is well known that Mercer's theorem can be applied to the Gaussian Schell-model source, and the cross-spectral density (19) can be expressed in the form [68]:

$$W^{(0)}(\mathbf{\rho}_1, \mathbf{\rho}_2) = \sum_{\ell=-\infty}^{+\infty} \sum_{n=0}^{+\infty} \Lambda_{n,\ell} F_{n,\ell}(\rho_1/\sigma_{\text{eff}}) F_{n,\ell}(\rho_2/\sigma_{\text{eff}}) \exp[-i\ell(\varphi_1 - \varphi_2)], \tag{21}$$

$$\Lambda_{n,\ell} = n!(2t)^{2n+|\ell|}(1-4t^2)/(n+|\ell|)!; \quad t = \Delta^2/2(1+\Delta^2+\sqrt{1+2\Delta^2}), \tag{22}$$

$$F_{n,\ell}(u) = u^{|\ell|} L_n^{|\ell|}(u^2) \exp(-u^2/2), \tag{23}$$

where $L_n^\ell(u)$ is a generalized Laguerre polynomial, $\sigma_{\text{eff}} = 1/(1+2\Delta^2)^{1/4}$ is a dimensionless effective mode width; and $F_{n,\ell}(u)\exp(i\ell\varphi)$ and $\Lambda_{n,\ell}$ are the eigenfunctions and eigenvalues of the integral equation, the kernel of which is the cross-spectral density [65]. Representation (21) can be interpreted as: Each term in the sum is the cross-spectral density for a fully coherent Laguerre–Gaussian mode. The eigenvalues, $\Lambda_{n,\ell}$, determine the "strength" that different modes contribute to the total energy of the entire beam [69]. At the same time, modes with different indices are mutually uncorrelated with each other.

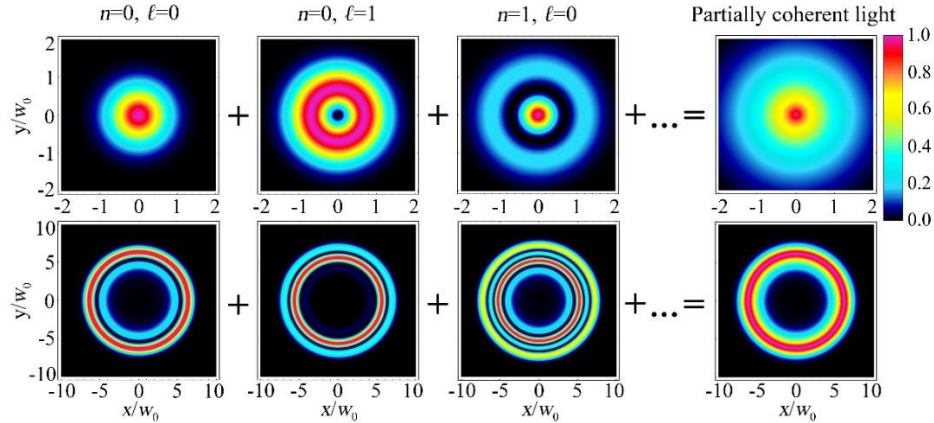

**Figure 2.** Illustration of the coherent-mode representation for a partially coherent light before (first row) and after the CR crystal (second row). The first column demonstrates the transverse intensity distribution for a fully coherent Gaussian beam ($n=0$, $\ell=0$). The second and third columns show Laguerre–Gaussian modes with indices $n=0$, $\ell=1$ and $n=1$, $\ell=0$. Their incoherent superposition with the other Laguerre–Gaussian modes leads to formation of a partially coherent radiation, as shown in the last column. Numerical simulations are run with $\rho_0=6$ and $\Delta=1.25$.

As a result, one can substitute coherent-mode representation of the Gaussian Schell-model source (21) into the expression for the orbital correlation functions (16), and, after simple algebra, the following expression can be obtained:

$$W_{\nu\mu}(\boldsymbol{\rho}_1,\boldsymbol{\rho}_2,\xi) = e^{-i\nu\varphi_1+i\mu\varphi_2} \sum_{\ell=-\infty}^{+\infty}\sum_{n=0}^{+\infty} \Lambda_{n,\ell} B_{\nu,n,\ell}(\rho_1,\xi)^* B_{\mu,n,\ell}(\rho_2,\xi) e^{-i\ell(\varphi_1-\varphi_2)}, \quad (24)$$

where $B_{\mu,n,\ell}(\rho,\xi)$ is the Belsky–Khapalyuk–Berry integral, which has already become classic [63,64]:

$$B_{\mu,n,\ell}(\rho,\xi) = \sigma_{\text{eff}}^2 \int_0^{+\infty} d\kappa\, \kappa\, F_{n,\ell}(\kappa\sigma_{\text{eff}}) \cos(\kappa\rho_0 - \mu\tfrac{\pi}{2}) e^{-i\kappa^2\xi/2} J_{\ell+\mu}(\kappa\rho), \quad (25)$$

where $J_m$ is the Bessel function of the first kind of order $m$. For a coinciding OAM ($\nu=\mu$), the orbital correlation function (24) is proportional to $\exp[i\mu(\varphi_2-\varphi_1)]$, meaning that the CR with partially coherent light can be expressed in terms of the coherence vortices [58] that have a similar vortex singularity in the cross-spectral density.

For the dual-cone model of the CR, the coherent-mode representation's form is similar to that of Equation (24):

$$V_{\nu\mu}^{(\sigma_1\sigma_2)}(\boldsymbol{\rho}_1,\boldsymbol{\rho}_2,\xi) = e^{-i\nu\varphi_1+i\mu\varphi_2} \sum_{\ell=-\infty}^{+\infty}\sum_{n=0}^{+\infty} \Lambda_{n,\ell} C_{\nu,n,\ell}^{(\sigma_1)}(\rho_1,\xi)^* C_{\mu,n,\ell}^{(\sigma_2)}(\rho_2,\xi) e^{-i\ell(\varphi_1-\varphi_2)}, \quad (26)$$

where $C^{(\sigma)}_{\mu,n,\ell}(\rho,\xi)$ is an integral of a dual-cone model [66,67]:

$$C_{\mu,n,\ell}^{(\pm)}(\rho,\xi) = \frac{1}{2}(\pm i)^\mu \sigma_{\text{eff}}^2 \int_0^{+\infty} d\kappa\, \kappa\, F_{n,\ell}(\kappa\sigma_{\text{eff}}) \exp(\mp i\kappa\rho_0) e^{-i\kappa^2\xi/2} J_{\ell+\mu}(\kappa\rho). \quad (27)$$

As a result, using the coherent-mode representation of the Gaussian Schell-model source, we demonstrate that the CR with partially coherent radiation can be expressed in terms of Belsky–Khapalyuk–Berry and dual-cone model integrals, as is clearly depicted in Figure 2. However, the coherent-mode representation itself does not provide an intuitive physical interpretation for a partially coherent CR. Furthermore, we develop an analytical approach that illuminates understanding of the physical properties associated with the aforementioned optical phenomenon.

## 4. Conical Refraction from Light With a Fluctuating Phase

Let us express the cross-spectral density (19) in terms of the coherent beams in which spatial evolution would be determined by a simple and understandable law, by analogy with the Laguerre–Gaussian representation (21). For this purpose, we use the property of the Schell-model source, rewritten as [70]:

$$W^{(0)}(\boldsymbol{\rho}_1,\boldsymbol{\rho}_2) = \int \frac{d\mathbf{q}}{2\pi}\, \tilde{g}(\mathbf{q})\, \mathcal{E}^*(\mathbf{q},\boldsymbol{\rho}_1)\mathcal{E}(\mathbf{q},\boldsymbol{\rho}_2), \quad (28)$$

where $\tilde{g}(\mathbf{q})$ is a Fourier transform of the degree of spatial coherence $g(\boldsymbol{\rho})$, $\mathbf{q}$ is a random transverse wave vector, and $\mathcal{E}(\mathbf{q},\boldsymbol{\rho})=\mathcal{E}^{(0)}(\boldsymbol{\rho})\exp[i\mathbf{q}\boldsymbol{\rho}]$ is a random electric field amplitude. Equation (28) can be physically interpreted as: Before the CR crystal, the electric field has a deterministic amplitude of $\mathcal{E}^{(0)}(\boldsymbol{\rho})$ but a fluctuating phase of $\Phi=\mathbf{q}\boldsymbol{\rho}$ [71]. Afterward, the ensemble averaging over the realizations of the electric field can then be understood as an integration over all possible random wave vectors $\mathbf{q}$ with a weight function of $\tilde{g}(\mathbf{q})$. The weight function directly contains all of the information on the coherence of the initial radiation.

As a result, representation (28) allows us to significantly simplify the calculation of the CR correlation functions. First, it is necessary to calculate the CR with an electric field with an amplitude $\mathcal{E}(\mathbf{q},\boldsymbol{\rho})$, depending on the random wave vector $\mathbf{q}$:

$$C_\mu^{(\pm)}(\mathbf{q},\boldsymbol{\rho},\xi) = \int \frac{d\boldsymbol{\kappa}}{2\pi} \exp(i\boldsymbol{\kappa}\boldsymbol{\rho})\, \tilde{G}_\mu^{(\pm)}(\boldsymbol{\kappa},\xi)\, \tilde{\mathcal{E}}^{(0)}(\boldsymbol{\kappa}-\mathbf{q}). \quad (29)$$

Then, from the calculated CR field amplitude (29), orbital correlation functions (14) can be obtained, after averaging over a random wave vector by analogy with (28).

$$V_{\nu\mu}^{(\sigma_1\sigma_2)}(\mathbf{\rho}_1,\mathbf{\rho}_2,\xi) = \int \frac{d\mathbf{q}}{2\pi} \tilde{g}(\mathbf{q}) C_\nu^{(\sigma_1)}(\mathbf{q},\mathbf{\rho},\xi)^* C_\mu^{(\sigma_2)}(\mathbf{q},\mathbf{\rho},\xi). \tag{30}$$

Let us consider the limit of high spatial coherence, with the weight function $\tilde{g}(\mathbf{q})$ localized near small wave vectors. Assuming that the longitudinal coordinate is $\xi=0$ since the case of small wave vectors is further used only in the focal plane, one can conveniently represent the components of the positive CR cone as: $C_\mu^{(+)}(\mathbf{q},\mathbf{\rho})=C_\mu^{(+)}(0,\mathbf{\rho})\exp[A(\mathbf{q},\mathbf{\rho})+iB(\mathbf{q},\mathbf{\rho})]$, where $A$ and $B$ are the amplitude and phase modulations emerging from the randomness of the transverse wave vector of the input beam. Both introduced modulations are given by the expressions:

$$A = \mathrm{Re}(ln[C_\mu^{(+)}(\mathbf{q},\mathbf{\rho})/C_\mu^{(+)}(0,\mathbf{\rho})]), \tag{31}$$

$$B = \mathrm{Im}(ln[C_\mu^{(+)}(\mathbf{q},\mathbf{\rho})/C_\mu^{(+)}(0,\mathbf{\rho})]). \tag{32}$$

The introduced representation is highly convenient since the difference between the exact cone component $C_\mu^{(\pm)}(\mathbf{q},\mathbf{\rho})$ and the unperturbed solution $C_\mu^{(\pm)}(0,\mathbf{\rho})$ is quite small. Therefore, amplitude and phase modulation can be considered as very small quantities. Worthy of mention is that both modulations are independent of the extra OAM $\mu$ because, in the limit of a well-distinguished CR ring ($\rho_0\gg 1$), it becomes insignificant. This property allows us to move from the components of the CR cone directly to the cone vector $\mathbf{C}^{(\pm)}$, given by formula (11):

$$\mathbf{C}^{(\pm)}(\mathbf{q},\mathbf{\rho}) = \mathbf{C}^{(\pm)}(0,\mathbf{\rho})\exp[A(\pm\mathbf{q},\mathbf{\rho})\pm iB(\pm\mathbf{q},\mathbf{\rho})], \tag{33}$$

where only two modulations, $A$ and $B$, are introduced for both cones, instead of four functions, $A^{(\pm)}$ and $B^{(\pm)}$, because we have the connection between the negative components of the CR cone and the positive one $C_\mu^{(-)}(\mathbf{q},\mathbf{\rho},\xi)=C_\mu^{(+)}(-\mathbf{q},\mathbf{\rho},-\xi)^*$. To derive this relation, we assume that the Fourier transform of the input field amplitude $\tilde{\mathscr{E}}^{(0)}(\mathbf{\kappa})$ is symmetric when $\mathbf{\kappa}$ is replaced by $-\mathbf{\kappa}$. As a result, now, only the amplitude and phase modulations depend on the random wave vector in Equation (33). Furthermore, it is demonstrated that the amplitude modulation shown in Figure 3a–b plays a crucial role compared to a phase modulation. Clearly, the amplitude modulation of the positive cone $\mathbf{C}^{(+)}$ is positive for $\rho_x>0$ and negative for $\rho_x<0$ (Figure 3a). Conversely, the amplitude modulation corresponding to the negative cone $\mathbf{C}^{(-)}$ behaves in the opposite way (Figure 3b). This is explained by the fact that the presence of a transverse wave vector leads to a tilt of the light beam direction by an angle proportional to the modulus of the wave vector. In this case, the increase in the wave vector transforms the CR into birefringence, as clearly demonstrated in Figure 3c–h [46,48,72]. Note that the positive cone corresponds to an extraordinary ray, and the negative cone corresponds to an ordinary ray.

At the same time, it is illogical to describe the intermediate intensity distributions in the form of a crescent (Figure 3e–f) and bright spots (Figure 3g–h) using amplitude and phase modulations (31)–(32), due to the large values of the random wave vectors. Let us directly calculate the components of the CR cone (29) for these cases. Because the deterministic amplitude of the electric field $\mathscr{E}^{(0)}(\mathbf{\rho})$ has a Gaussian form (20), we use the Jacobi–Anger expansion [73] and integrate Equation (29) over the angle variable, obtaining:

$$C_\mu^{(+)}(\mathbf{q},\mathbf{\rho},\xi) = \frac{e^{-\frac{q^2}{2}}}{2} \sum_{m=-\infty}^{+\infty} i^\mu e^{i(m+\mu)\varphi-im\theta_q} \int_0^{+\infty} \kappa d\kappa\, e^{-\frac{(1+i\xi)\kappa^2}{2}-i\rho_0\kappa} J_m(iq\kappa) J_{m+\mu}(\kappa\rho). \tag{34}$$

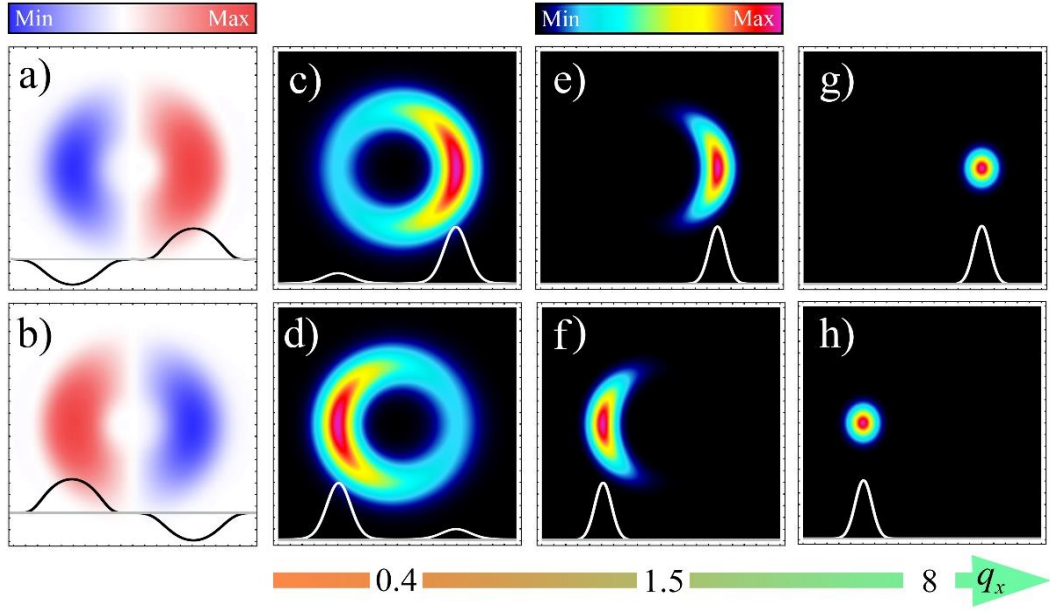

**Figure 3.** (a)–(b) Transverse distributions of amplitude modulation of the CR cones in the focal plane. For the numerical calculation, we use Equation (31) and the parameters $\rho_0=5$ and $q_x=0.4$. (c)–(h) Transverse intensity distributions of the CR cones in the focal plane are calculated with Equation (11) and (29) for the wave vectors $q_x=0.4$ (c)–(d), $q_x=1.5$ (e)–(f), and $q_x=8$ (g)–(h). The first row corresponds to the positive cone $\mathbf{C}^{(+)}$, and the second to the negative cone $\mathbf{C}^{(-)}$. For clarity, associated central cross-sections along the x-axis are shown at the bottom of each figure.

Equation (34) clearly demonstrates that, in the limit of large wave vectors, the CR beam can be represented as a superposition of Bessel-Gaussian modes with a different OAM passing through the CR crystal [74]. Thus, if the transverse wave vector is small, the term with a zero OAM then makes the main contribution to the sum, and we obtain the CR of a Gaussian beam. For an arbitrary wave vector, higher OAMs also become important. Thus, an increase in the transverse wave vector increases the contribution of higher OAMs, leading to the transformation of the CR cones into a crescent-shaped beam (Figure 3e–f) and then into a bright spot (Figure 3g–h).

It may seem that expression (34) is final because the CR integrals cannot be explicitly calculated [75,76]. However, we can apply the condition of large wave vectors ($|\mathbf{q}|\gg 1$), as well as the Bessel-Gaussian model of the CR [53], transforming expression (34) into the following form (Appendix 2):

$$C_\mu^{(+)}(\mathbf{q},\boldsymbol{\rho},\xi) = \frac{\sqrt{-2\pi i \varepsilon \tau^2}e^{i\mu\varphi}}{2\sigma_\xi}\left(\frac{\rho+\sigma_\xi\rho_0\varepsilon^2 e^{-i(\varphi-\theta_q)}}{\rho+\sigma_\xi\rho_0\varepsilon^2 e^{i(\varphi-\theta_q)}}\right)^{\frac{\mu}{2}} \times$$
$$\times \exp\left[-\frac{\rho^2+\rho_\xi^2}{2\sigma_\xi} - i\mathbf{q}\boldsymbol{\rho}_0 - i\xi\frac{q^2}{2} - \tau\rho_0\varepsilon^2\right] I_\mu\left(\frac{\tau}{\sigma_\xi}\sqrt{\rho^2+2\sigma_\xi\boldsymbol{\rho}\rho_0\varepsilon^2+\sigma_\xi^2\rho_0^2\varepsilon^4}\right), \quad (35)$$

where $I_m(x)$ is the modified Bessel function of the first kind of order $m$, and $\tau=\rho_0+iq$ is a complex propagation parameter introduced to describe generalized Bessel-Gaussian beams [77] (the real part of propagation parameter $\tau$ specifies the beam radius in the focal plane $\rho_0$, and the imaginary part determines the tilt angle $q$, with respect to the beam symmetry axis). In addition, $\boldsymbol{\rho}_\xi=\rho_\xi\mathbf{n}_q$ is a vector, wherein its modulus determines a ring radius for an arbitrary $\xi$ value and is equal to $\rho_\xi=\rho_0+q\xi$, and its direction is determined by the normalized transverse wave vector $\mathbf{n}_q=\mathbf{q}/q$; $\sigma_\xi=1+i\xi$ is a spatial propagation function; and $\varepsilon=q/\rho_0$ is a beam asymmetry parameter that determines the transition from the crescent-shaped beam ($\varepsilon\ll 1$), shown in Figure 3e–f, to the bright spots ($\varepsilon\gg 1$), shown in Figure 3g–h. Thus, if the asymmetry parameter is very large ($\varepsilon\gg 1$), then, Equation (35) is greatly simplified to a Gaussian beam:

$$C_\mu^{(+)}(\mathbf{q},\boldsymbol{\rho},\xi) = \frac{e^{i\mu\theta_q}}{2\sigma_\xi}\exp\left[-\frac{(\boldsymbol{\rho}-\boldsymbol{\rho}_\xi)^2}{2\sigma_\xi} + i\mathbf{q}(\boldsymbol{\rho}-\boldsymbol{\rho}_0) - i\xi\frac{q^2}{2}\right]. \quad (36)$$

Worthy of note is that the introduced beams (35) are very similar to the asymmetric Bessel [78,79] and Bessel-Gaussian [80,81] beams previously described in the literature. The main difference is that the propagation parameter τ is now a complex variable, just as for generalized Bessel-Gaussian beams [77]. Hence, the introduced beams should be called generalized asymmetric Bessel-Gaussian (GABG) beams.

In this section, we demonstrate the formalism by describing the partially coherent CR, as well as using a representation in which the input electric field has a deterministic amplitude but a fluctuating phase. Here, we can still calculate the propagation through the CR crystal, and average fluctuations, only at the last stage. For the Schell-model source, the fluctuating phase is expressed in terms of a random wave vector. When the random wave vector is small, it is convenient to describe the CR beam in terms of amplitude and phase modulations (33). In the intermediate case, when the modulus of the wave vector $q$ is much greater than unity but much less than the normalized CR ring radius $\rho_0$, the intensity distribution is transformed into the crescent-shaped GABG beam (35). In the limit of quite large wave vectors ($q \gg \rho_0$), the CR completely transforms into birefringence and is expressed in terms of the usual Gaussian beams (36). Furthermore, we will use the exact solutions obtained in this section to calculate the CR intensity for the near and far field regions.

## 5. CR Rings in the Focal Plane

In this section, we consider a partially coherent CR in the focal plane. Previously, it was demonstrated [62] that a decrease in the spatial coherence of light leads to a disappearance of the classical double-ring Lloyd's intensity distribution, as shown in Figure 4a. To explain this effect, we derive a phenomenological theory [62] based on the dual-cone model of the CR [66,67]. Within its framework, the CR intensity for coherent radiation is expressed in terms of the CR cones as $I = |\mathbf{C}^{(+)} + \mathbf{C}^{(-)}|^2$. In accordance with phenomenological theory [62], the partial spatial coherence of light leads to the following modification of the intensity in the focal plane ($\xi = 0$):

$$I(\boldsymbol{\rho}) = |\mathbf{C}^{(+)}(\boldsymbol{\rho})|^2 + |\mathbf{C}^{(-)}(\boldsymbol{\rho})|^2 + 2|\mathbf{C}^{(+)}(\boldsymbol{\rho})||\mathbf{C}^{(-)}(\boldsymbol{\rho})|\alpha_{CR}\cos\left[\arg(\mathbf{C}^{(-)}(\boldsymbol{\rho})^*\mathbf{C}^{(+)}(\boldsymbol{\rho}))\right], \qquad (37)$$

where $\alpha_{CR}$ is a phenomenological coherence degree of the CR cones. When the coherence degree is $\alpha_{CR}=1$, the radiation is completely coherent, and the CR cones interfere with each other, producing a double-ring intensity distribution. When the radiation is incoherent, and the coherence degree is $\alpha_{CR}=0$, the interference term in Equation (37) vanishes, and we observe complete disappearance of the dark Poggendorff's ring in the focal plane. If the radiation is partially coherent, then, the degree of coherence $\alpha_{CR}$ takes values between zero and unity. At the same time, the relationship between the phenomenological coherence degree of the CR cones and the input spatial degree of coherence has the following simple form: $\alpha_{CR}=g(\boldsymbol{\rho}_C)$, where $\boldsymbol{\rho}_C$ is a characteristic distance, depending on the intensity distribution of the input radiation [62].

Furthermore, we will show that the phenomenological model of partially coherent CR can be rigorously justified for the case of high spatial coherence, when the coherence parameter is $\Delta \ll 1$. We prove the relation $\alpha_{CR}=g(\boldsymbol{\rho}_C)$ and connect the characteristic distance $\boldsymbol{\rho}_C$ with the properties of the input intensity distribution. Furthermore, we obtain a physical explanation for the coherence degree of the CR cones, $\alpha_{CR}$. In the low-coherent regime ($\Delta \gg 1$), the coherence degree $\alpha_{CR}$ decreases according to a power law, which, up to a constant, does not depend on the detailed form of the input spatial degree of coherence. In addition, we predict a counterintuitive effect, or narrowing of the CR ring, with a reduction of the light coherence.

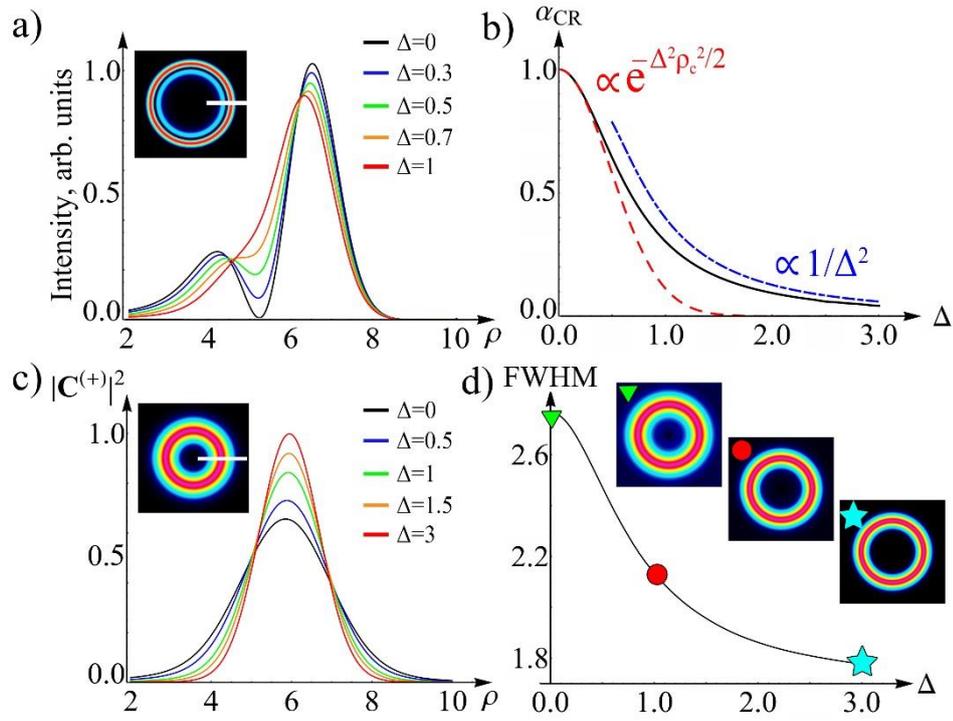

**Figure 4.** (a) Transition from a spatially coherent light-intensity profile of the CR beam in the Lloyd's plane (low Δ) to a low-coherent one (high Δ). (b) The dependence of the CR cones coherence degree $\alpha_{CR}$ on the coherence parameter Δ. The calculation was carried with the dual-cone integrals (26) (black), a high-coherent approximation (43) (red), and a low-coherent one (47) (blue). (c) Light-intensity profile for the positive CR cone in the Lloyd's plane for different Δ coherence parameters. (d) The CR cone full width at half maximum (FWHM) versus the coherence parameter Δ. Associated transverse intensity distributions are shown at the top of (a),(c),(d). The normalized CR radius is set to $\rho_0=6$.

First, we introduce a strict definition for the phenomenological coherence degree of the CR cones, using unified optical coherence theory. This can be done if the radiation intensity is expressed in terms of the CR cones by analogy with the phenomenological Equation (37):

$$I(\boldsymbol{\rho}) = I^{(+)}(\boldsymbol{\rho}) + I^{(-)}(\boldsymbol{\rho}) + 2\left[I^{(+)}(\boldsymbol{\rho})\right]^{1/2}\left[I^{(-)}(\boldsymbol{\rho})\right]^{1/2}|\gamma(\boldsymbol{\rho},\boldsymbol{\rho})|\cos\left[\arg\gamma(\boldsymbol{\rho},\boldsymbol{\rho})\right], \qquad (38)$$

where $I^{(\pm)}(\boldsymbol{\rho})=<|\mathbf{C}^{(\pm)}(\boldsymbol{\rho})|^2>$ is an average intensity of each of the cones; and $\gamma(\boldsymbol{\rho}_1,\boldsymbol{\rho}_2)=<\mathbf{C}^{(-)}(\boldsymbol{\rho}_1)^*\mathbf{C}^{(+)}(\boldsymbol{\rho}_2)>/I^{(+)}(\boldsymbol{\rho}_1)^{1/2}I^{(-)}(\boldsymbol{\rho}_2)^{1/2}$ is a normalized cross-cone correlation function, introduced by analogy with the classical problem of the interference of light passing through a screen with two parallel slits [65]. Therefore, from formulas (37) and (38), the modulus of the normalized cross-cone correlation function in the exact theory acts as the coherence degree of the CR cones:

$$\alpha_{CR} = |\gamma(\boldsymbol{\rho},\boldsymbol{\rho})| = \left|\left\langle \mathbf{C}^{(-)}(\boldsymbol{\rho})^*\mathbf{C}^{(+)}(\boldsymbol{\rho})\right\rangle\right|/\sqrt{I^{(+)}(\boldsymbol{\rho})}\sqrt{I^{(-)}(\boldsymbol{\rho})}. \qquad (39)$$

Generally, the coherence degree of the CR cones, $\alpha_{CR}$, depends on the transverse radius vector **ρ**. However, we will consider the area near the maximum intensity of the CR ring, and, therefore, $\rho \approx \rho_0$. Formula (39) demonstrates that, in order to obtain the degree of coherence $\alpha_{CR}$, one must first calculate the intensity of the cones and the cross-cone correlation function. To this end, we use the representation of the fluctuating phase (30). This rigorous calculation for the high spatial coherence regime and the Schell-model source is performed in Appendix 3, resulting in the following expressions:

$$\left|\left\langle \mathbf{C}^{(-)}(\boldsymbol{\rho})^*\mathbf{C}^{(+)}(\boldsymbol{\rho})\right\rangle\right| \propto \left|\mathbf{C}^{(-)}(0,\boldsymbol{\rho})\right|\left|\mathbf{C}^{(+)}(0,\boldsymbol{\rho})\right|, \qquad (40)$$

$$I^{(\pm)}(\boldsymbol{\rho}) \propto \left|\mathbf{C}^{(\pm)}(0,\boldsymbol{\rho})\right|^2 / g(\boldsymbol{\rho}_C), \qquad (41)$$

where $\mathbf{C}^{(\pm)}(0,\boldsymbol{\rho})$ is the unperturbed cone, and $\boldsymbol{\rho}_C$ is a characteristic spatial distance, expressed in terms of the Fourier transform of the deterministic field amplitude $\tilde{\mathscr{E}}^{(0)}(\boldsymbol{\kappa})$, as:

$$\boldsymbol{\rho}_{\mathrm{C}} = \mathbf{e}_r \int_0^{+\infty} d\kappa \, \kappa^{-1/2} \, \tilde{\mathcal{E}}^{(0)}(\kappa) \Big/ \int_0^{+\infty} d\kappa \, \kappa^{1/2} \, \tilde{\mathcal{E}}^{(0)}(\kappa), \qquad (42)$$

where $\mathbf{e}_r=[\cos(\varphi), \sin(\varphi)]$ is a radial unit vector in a polar coordinate system. As a result, using expressions (39)–(41), it is possible to express the coherence degree of the CR cones $\alpha_{CR}$, in terms of the input spatial degree of coherence, in the following simple, elegant way:

$$\alpha_{CR} = g(\boldsymbol{\rho}_{\mathrm{C}}). \qquad (43)$$

Thus, we confirm the relationship between the input and output degrees of coherence, phenomenologically obtained earlier [62], for the high spatial coherence limit. From Equation (43), it is clear that a decrease in the input beam coherence causes the disappearance of the correlation between two CR cones, leading to the disappearance of the interference pattern, shown in Figure 4a. In the search for physical explanations, it may be tempting to explain the behavior of coherence degree of the CR cones by reducing the numerator in Equation (39) with a fixed denominator. Such a statement is typical in the theory of optical coherence and is associated with phase fluctuation that appears in the cross-correlation function but disappears in the intensity. However, analysis of Equations (40)–(41) leads to the opposite conclusion. Thus, the cross-cone correlation function does not change, in contrast to the intensity, wherein the input spatial degree of coherence appears in the denominator, because of the critical role of amplitude, rather than phase fluctuations. This observation is underscored by the transverse distributions for the amplitude modulation of the positive and negative cones, shown in Figure 3a–b. From this figure, it is clear that the product of two different cones, determining the cross-cone correlation function (40), does not depend on the random wave vector and, hence, on the input light coherence because the amplitude modulations cancel each other out. However, in the case of two identical cones, determining the cone intensities (41), the modulations are doubled; therefore, the intensity acquires a strong dependence on the input light coherence. Thus, the amplitude fluctuations of the CR field are the physical reason for the appearance of the coherence degree, $\alpha_{CR}$. Comparison of the coherence degree of the CR cones, obtained from Equation (43), and the numerical simulation, using the coherent-mode representation, is shown in Figure 4b. Here, good agreement is demonstrated between the analytical and numerical results for the case of high spatial coherence ($\Delta \ll 1$).

Now, we consider the case of low spatial coherence. The main difference here, from the case of high coherence is that the CR beam, before averaging, is now expressed in terms of GABG beams (35). It turns out that a simpler expression with Gaussian beams (36) can be used to calculate the intensity of the cones, resulting in:

$$I^{(\pm)}(\boldsymbol{\rho}) = \frac{1}{2} \int_0^{2\pi} d\theta_q \exp\left[-(\boldsymbol{\rho} - \rho_0 \mathbf{n}_q)^2\right] = \frac{1}{2} \exp\left[-\rho^2 - \rho_0^2\right] I_0(2\rho\rho_0), \qquad (44)$$

where the cone intensity is clearly independent of the input spatial degree of coherence. Using the approximation of a well-defined CR ring ($\rho_0 \gg 1$), we replace the modified Bessel function by its asymptotics, obtaining the intensity of the cones described by the Gaussian function $I^{(\pm)}(\boldsymbol{\rho}) \propto \exp[-(\rho-\rho_0)^2]$, a fact directly understood from analysis of formula (44). The resulting intensity is formed by an incoherent superposition of Gaussian beams, with its centers located on a ring of radius $\rho_0$. Averaging over the Gaussian beam centers results in the annular intensity distribution having a Gaussian profile. It is surprising, however, that the waist of the ring for incoherent radiation, defined by expression (44), will be smaller than that for the completely coherent CR ring, as demonstrated in Figure 4c.. This counterintuitive effect is explained by the behavior of the CR ring width for fully coherent light. Since the angular spectrum of the coherent CR ring is stronger localized than the spectrum of the initial Gaussian beam [53], the full width at half maximum (FWHM) of the ring will be greater than the FWHM of the input Gaussian beam. However, for incoherent radiation, the FWHM of the CR ring coincides with the FWHM of the original beam, as previously shown. Accordingly, a decrease in coherence leads to an effective decrease in the FWHM of the ring, as clearly demonstrated in Figure 4d.

To obtain the coherence degree of the CR cones, $\alpha_{CR}$, for the low-coherent regime, it is necessary to calculate the cross-cone correlation function using Equations (30) and (35). We show that the coherence degree of the CR cones depends on the input spatial degree of coherence, according to the following law (Appendix 3):

$$\alpha_{CR} = \int_0^{+\infty} d\rho\, \rho\, g(\rho) f(\rho),\qquad(45)$$

where $f(\rho)=\pi\exp[-\rho^2/16]M_{1,0}(\rho^2/8)/2$, and $M_{k,\mu}(x)$ is a Whittaker function. In the limit of low spatial coherence ($\Delta\gg 1$), the function $g(\rho)$ is localized on much smaller scales of the order of $1/\Delta$ than the function $f(\rho)$. Therefore, in the expression under the integral sign, the function $f(\rho)$ can be replaced by $f(0)=\pi/4\sqrt{2}$. As a result, the coherence degree of the CR cones is transformed into:

$$\alpha_{CR} \approx \frac{\pi}{4\sqrt{2}} \int_0^{+\infty} d\rho\, \rho\, g(\rho) = \frac{\pi}{4\sqrt{2}}\tilde{g}(0).\qquad(46)$$

The dependence of coherence degree $\alpha_{CR}$ on the coherence parameter $\Delta$ is easily obtained from (46). As aforementioned, $g(\rho)$ is localized on scales $1/\Delta$; therefore, one can change the integration variable to $u=\Delta\rho$. Here, the integral $\int du\, u\, g(u/\Delta)$ no longer depends on the coherence parameter $\Delta$, and we obtain the universal power law behavior $\alpha_{CR}\propto 1/\Delta^2 \propto (w_g/w_0)^2$. In the case of a Gaussian-Schell-model source, the coherence degree $\alpha_{CR}$ is easily calculated using the exact formula (45):

$$\alpha_{CR} = \frac{\pi\sqrt{2}}{(4+1/\Delta^2)^{3/2}}\frac{1}{\Delta^2},\qquad(47)$$

confirming that, for the low coherence limit ($\Delta\gg 1$), the coherence degree of the CR cones decreases, according to the power law.

This behavior of the coherence degree can be explained in a following way: As aforementioned, the cone intensity (44) does not depend on the coherence parameter $\Delta$. Therefore, the main contribution to the coherence degree is demonstrated by the cross-cone correlation function obtained via the averaging over random wave vectors (30). Resultingly, the largest contribution to the correlation function is made by random wave vectors near the ring with radius $q\approx 1$-$2$. The area of this region in **q**-space is $S_0 \propto 1$. However, the weight function $\tilde{g}(\mathbf{q})$ is not equal to zero in the region $S_{All} \propto \Delta^2$. Thus, the smaller the degree of coherence, and the larger coherence parameter $\Delta$, the smaller the ratio of wave vectors that contribute to the coherence degree $\alpha_{CR} \propto S_0/S_{All} \propto 1/\Delta^2 \propto (w_g/w_0)^2$. In conclusion, we find a universal power law dependence of the degree of coherence of the CR cones $\alpha_{CR}$ versus the correlation length $w_g$. Figure 4 illustrates the resulting power law dependence in agreement with the numerical calculation of the coherence degree $\alpha_{CR}$ for the Gaussian-Schell model source. The figure also clearly shows how the transition occurs from the high-coherent regime (43) to low-coherent one (47).

As a result, in this section, we have described the behavior of the coherence degree of the CR cones in the limit of high and low spatial coherence, as well as demonstrated the counterintuitive effect of narrowing of the CR ring with reduction of the light coherence. In the next section, we will consider how the far field of a CR beam depends on the input spatial coherence.

## 6. Axial Spike

Let us consider the influence of the light coherence on the formation of the axial spike near the propagation axis ($\rho\approx 0$). For that purpose, one can study the orbital correlation functions $W_{\mu\mu}(\boldsymbol{\rho},\boldsymbol{\rho},\xi)$. Equation (10) demonstrates that $W_{\mu\mu}(\boldsymbol{\rho},\boldsymbol{\rho},\xi)$ can be interpreted as the light intensity with a given additional OAM equal to $\mu$. The total CR intensity is obtained after summing all orbital functions $W_{\mu\mu}(\boldsymbol{\rho},\boldsymbol{\rho},\xi)$ over the extra OAM $\mu$. To calculate such orbital correlation functions in the far zone ($\xi > 1$), one can directly use the multidimensional integral (16). Using the stationary phase method, we integrate Equation (16) over all radial variables, obtaining:

$$W_{\mu\mu}(\boldsymbol{\rho},\boldsymbol{\rho},\xi) = \frac{\pi\rho_0^2}{2(1+2\Delta^2)\xi^3} e^{-\frac{1}{1+2\Delta^2}\frac{\rho_0^2}{\xi^2}} \int_0^{2\pi}\frac{d\theta_1}{2\pi}\int_0^{2\pi}\frac{d\theta_2}{2\pi} \exp\left[-\frac{(\mathbf{n}_2-\mathbf{n}_1)^2}{2\theta_c^2}\right]\exp\left[i\frac{\rho_0}{\xi}(\mathbf{n}_2-\mathbf{n}_1)\boldsymbol{\rho}+i\mu(\theta_2-\theta_1)\right],\quad(48)$$

where $\mathbf{n}_{1,2}=[\cos(\theta_{1,2}),\sin(\theta_{1,2})]$ are the normal vectors, and $\theta_c=\xi(2+1/\Delta)^{1/2}/\rho_0$ is an correlation angle. For the physical meaning of Equation (48), one can consider the product of two amplitudes of plane waves exp[-$i\kappa_1\rho$] and exp[$i\kappa_2\rho$] with wave vectors $\kappa_{1,2}= \rho_0\mathbf{n}_{1,2}/\xi$ under the integral sign. After averaging this product over all possible directions of the normal vectors with the weight function exp[-$(\mathbf{n}_2-\mathbf{n}_1)^2/2\theta_c^2$], we obtain the resulting orbital correlation function. When the input radiation is coherent ($\Delta\ll 1$), the correlation angle is large, i.e., $\theta_c\propto\Delta^{-1/2}\gg 1$. As a result, the weight function exp[-$(\mathbf{n}_2-\mathbf{n}_1)^2/2\theta_c^2$]$\approx 1$ and the plane waves in expression (48) correlate with each other. In this case, the integrals over angles can be easily calculated, and we obtain the well-known expression for orbital correlation functions, $W_{\mu\mu}(\rho,\rho,\xi) \propto J_\mu(\rho\rho_0/\xi)^2$ [64], i.e., for coherent light near the axis in the far zone, the Raman spot is formed from several Bessel beams with an extra OAM $\mu=0,\pm1$. A similar result is also valid for low-coherent light ($\Delta\gg 1$) in the spatial region, where axial distance is greater than the normalized CR radius ($\xi>\rho_0$). This is due to the fact that the correlation angle in this region is also much larger than unity ($\theta_c\propto\xi/\rho_0>1$), as it is for coherent light. This statement is consistent with the Van Cittert–Zernike theorem [65], which stipulates that the transverse radius of light coherence increases with distance from the source. However, if the axial distance is less than the normalized CR radius ($\xi<\rho_0$), then, given low-coherent light, the correlation angle $\theta_c$ is much less than unity ($\theta_c\propto\xi/\rho_0<1$). In this case, the orbital correlation function (48) is easily calculated, and we can conclude:

$$W_{\mu\mu}(\rho,\rho,\xi) = \frac{\sqrt{\pi}\rho_0}{4\Delta^2\xi^2}\exp\left[-\frac{1}{2\Delta^2}\frac{\rho_0^2}{\xi^2}\right]\exp\left[-\frac{\rho^2}{2}\right]I_0\left(\frac{\rho^2}{2}\right). \quad (49)$$

In expression (49), the function exp[-$\rho_0^2/2\Delta^2\xi^2$]/$\xi^2$ is responsible for the axial evolution, wherein the maximum is located at the point $\xi_{max}=\rho_0/\Delta\sqrt{2}$, i.e., an increase in the coherence parameter $\Delta$, corresponding to a decrease in beam coherence, leads to a shift of the Raman spot closer to the focal plane, as shown in Figure 5a–b. The physical meaning of this effect can be found in the representation of partially coherent light via a fluctuating phase (30). Obviously, a decrease in coherence, due to an increase in the coherence parameter $\Delta$, leads to an increase in the characteristic random wave vector of the beam. In addition, the larger the transverse wave vector, the larger the angle at which the CR field forms an axial spike near the propagation axis. As a result, the Raman spots will move increasingly closer to the focal plane, with a decreasing input beam coherence. However, the most surprising phenomenon in expression (49) is a radial intensity distribution, given by the function exp[-$\rho^2/2$]$I_0(\rho^2/2)$. Since this transverse distribution does not depend on the axial distance $\xi$, we can conclude that the CR beam propagates without a diffractive spread, a counterintuitive result because the opposite effect can be observed for the Gaussian Schell-model source. It is well known that the divergence of such a beam increases with decreasing spatial coherence.

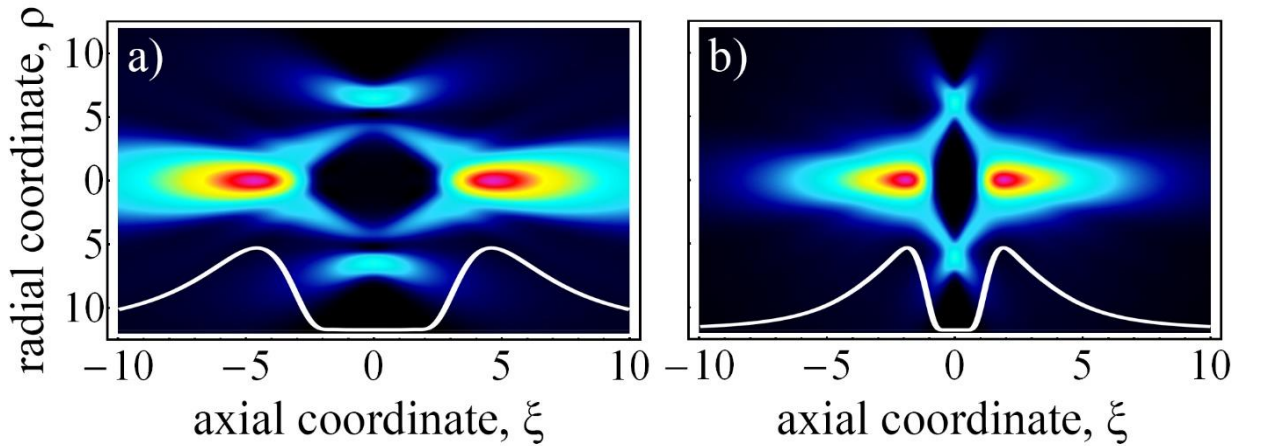

**Figure 5.** The axial distributions of the CR beam intensity are computed numerically from the mode decomposition (26), for (a) high- and (b) low-coherent light, with $\Delta=0$ and $\Delta=2$, correspondingly. For clarity, associated axial intensity distributions are shown at the bottom of each figure, as white lines.

The occurrence of a diffraction-free beam can be explained by decomposition of Equation (49), through the Bessel beams, with simple spatial evolution:

$$W_{\mu\mu}(\rho,\rho,\xi) \propto \sum_{m=-\infty}^{+\infty} J_{m-\mu}(\rho\rho_0/\xi)^2 \exp\left[-m^2\xi^2/\rho_0^2\right], \qquad (50)$$

where each Bessel beam undergoes diffractive spreading. The weight with which each Bessel beam enters the final expression is given by an exponential function, $\exp[-m^2\xi^2/\rho_0^2]$. Moreover, the weight function also depends on the axial distance $\xi$. Using formula (50), an estimate for the half-width of the orbital correlation function is easily obtained. Thus, the half-width of the mode with index $m$, at a distance $\xi$ from the focal plane, is proportional to $w \approx m\xi/\rho_0$, for $m \gg 1$, i.e., the larger the axial distance $\xi$, or the mode index $m$, the larger the half-width of the Bessel beam. However, the number of modes involved in the formation of the orbital correlation function distribution is limited by the weight function $\exp[-m^2\xi^2/\rho_0^2]$ and is equal to $m_{\max} \approx \rho_0/\xi$. Thus, the maximum number of modes will decrease with increasing axial distance $\xi$. Therefore, substituting the maximum mode index $m_{\max}$ into the mode half-width $w$, we obtain $w \propto 1$. As expected, the dependence of the half-width of the CR beam on the distance $\xi$ disappears, given that the effect of diffraction broadening is compensated by a multimode structure of partially coherent radiation.

## 7. Discussion

Three main effects of partially coherent CR that can be observed in the focal plane are represented in Figure 4. The first effect is the disappearance of the dark Poggendorff's ring in the Lloyd's distribution with reduction of the input beam coherence, as shown in Figure 4a. In this figure we demonstrate the numerically calculated intensity for different values of the input spatial coherence using developed cross-spectral density formalism discussed in the first and second sections. Here one can see transformation of a two-ring pattern into a bright annular ring with a decreasing coherence. The explanation of this effect lies in the interference nature of the double-ring intensity distribution. It arises when two CR cones intersect each other in the focal plane, forming an interference pattern, as shown in Figure 1b. This situation changes dramatically for partially coherent light, with vanishing interference. We reported on this phenomenon in our previous work [62], and for its theoretical description we phenomenologically introduce the coherence degree of CR cones [62], emerging in front of interference contribution of CR intensity distribution (37). It appears that the given phenomenological concept characterizes the essential properties of the effect under discussion. However, a deep understanding of the underlying connections between the CR light transformation and coherence of the incident radiation is out of consideration of the phenomenological theory. To fill the gap in understanding the internal mechanism of this effect, in this paper we study the dependence of the coherence degree of CR cones on the input light coherence, which is depicted in Figure 4b. We did it both numerically and analytically, using a rigorous definition of coherence degree in Equation (38). In Figure 4b, two characteristic regions of high and low spatial coherence can be distinguished, which are characterized by low and large coherence parameter $\Delta$, correspondingly. The highly coherent approximation is described by Equation (43), which is a simple and elegant relationship between the output and input spatial degrees of coherence that was introduced phenomenologically [62] and confirmed here from the first principles. To obtain Equation (43) and corresponding physical explanation for the CR coherence degree we use phase fluctuation representation of the initial cross-spectral density [70], which was discussed in the third section. The main idea of the applied formalism is that the electric field vector has a deterministic amplitude, but a fluctuating phase, which is expressed in terms of random transverse wave vector. It turns out that the phase fluctuations of the input field generate amplitude modulation of the CR field as shown in Figure 3a-d. This modulation has an opposite behavior for different CR cones, which significantly affects the intensity of both CR cones and does not disturb the cross-cone correlation function. As a result, the coherence degree of CR cones emerges in front of the interference contribution to the CR intensity in Equation (38).

The second new effect that can be observed in the focal plane is a universal power-law dependence of the coherence degree of CR cones on input correlation length, which is described by equations (45)-(47) and is clearly seen in Figure 4b. In this low-coherent regime, the CR coherence degree is independent of the detailed form of the input spatial coherence degree, up to a constant before the power-law. For analytical calculations of this result we also use the phase fluctuation representation, but in the limit of high random transverse wave vectors. For this regime, the CR cones before averaging are described by Equation (35) and the introduced generalized asymmetric Bessel-Gaussian beams, which have a crescent-like shape, as shown in Figure 3e-f. Their behavior in the region of critical wave vectors separating the low and high coherence regimes completely determines the predicted universal power-law, as shown in Appendix 3.

Furthermore, it is interesting to consider the transformation of the CR cones intensities themselves from high to low coherence, which is depicted in Figure 4c. Surprisingly, the width of the CR ring becomes narrower with reduction of the spatial coherence degree. This phenomenon is the third new effect occurring in the focal plane. It can be easily explained by considering the coherent and incoherent limits for the FWHM of the CR ring, the dependence of which on the input coherence is shown in Figure 4d. In coherent case, FWHM is greater than that of the incident beam due to the strong angular localization of the beam behind the CR crystal. For the incoherent radiation, CR intensity in the focal plane consists of superposition of the Gaussian beams shown in Figure 3g-h, whose centers are randomly placed on the CR ring. Thus, the FWHM of the resulting ring coincides with the incident one. As a result, we will observe an effective decrease in the FWHM of the ring with decreasing coherence.

After focal plane we focused our attention on the far field zone. Thus, the calculated CR axial evolution for coherent and incoherent radiation is depicted in Figure 5. On figure 5b, one can observe the formation of the diffraction-free beam and the Raman spot shift. The first effect can be derived from the analytically calculated orbital correlation function described by Equation (49) and is due to the multimode structure of low-coherent radiation that suppresses diffraction broadening. The second phenomenon emerges from the spatial evolution of a conically refracted Gaussian beam with a nonzero random transverse wave vector. As the input coherence degree decreases, the characteristic wave vector increases. Since the angle at which the CR beam propagates with respect to the optical axis is proportional to the wave vector, the axial spike of the CR beam will shift towards the focal plane as the input coherence decreases. All new predicted low coherent effects are out of consideration of the phenomenological theory and are very nonstandard in comparison with the properties of the Gaussian Schell-model source.

## 8. Conclusion

In conclusion, we derive a theory of partially coherent conical refraction, which is in agreement with the available experiments. Then, we demonstrate that the CR with partially coherent radiation can be expressed in terms of Belsky–Khapalyuk–Berry and dual-cone model integrals. We, also, reformulated and significantly simplified the rigorous CR theory, assuming randomness of the electrical field phase of the input beam. It allows us to calculate the propagation through the CR crystal, and average fluctuations, only at the last stage. We use the obtained exact solutions to calculate the CR intensity for the near and far field regions. As a result, we explain disappearance of the dark Poggendorff's ring in the Lloyd's plane and predict counterintuitive effect of narrowing of the CR ring width and shift of Raman spots for the low-coherent CR light. We also demonstrate a universal power-law dependence of CR cones coherence degree on the input correlation length and diffraction-free propagation of the low-coherent CR light in the far field.

## 9. Appendix 1: Polarization Structure of the CR

Let us find the relationship between the cross-spectral density matrix (7) and orbital correlation functions (8). Here, using Equation (1) for the CR field, we transform the expression for the cross-spectral density matrix (7) into the following form:

$$K_{ij}(\mathbf{\rho}_1,\mathbf{\rho}_2,\xi) = \sum_{k,s=R,L} \left\langle \beta_{ik}^*(\mathbf{\rho}_1,\xi)\beta_{js}(\mathbf{\rho}_2,\xi) \right\rangle J_{ks}, \qquad (51)$$

where we introduce the matrix $\beta$:

$$\hat{\beta}(\mathbf{\rho},\xi) = \begin{pmatrix} \mathcal{B}_0(\mathbf{\rho},\xi) & \mathcal{B}_{-1}(\mathbf{\rho},\xi) \\ \mathcal{B}_1(\mathbf{\rho},\xi) & \mathcal{B}_0(\mathbf{\rho},\xi) \end{pmatrix}, \qquad (52)$$

with its elements being the components of the CR field (2). From Equation (52), it is clear that each $\langle \beta_{ik}^*\beta_{js}\rangle$ component is associated with the corresponding orbital correlation function, $\langle B_\nu^* B_\mu \rangle$. Therefore, using formulas (51)–(52), we obtain relation (9), connecting the cross-spectral density matrix (7), with the introduced orbital correlation functions (8) and the polarization matrix $J_{ij}$.

Knowing relation (9), one can easily find the relationship between the CR intensity and the orbital correlation functions, $W_{\mu\nu}$. To obtain this relationship, one can rewrite cross-spectral density and polarization matrices in terms of the Stokes parameters [65].

$$\hat{K}(\mathbf{\rho},\mathbf{\rho},\xi) = \frac{1}{2} \begin{pmatrix} S_0(\mathbf{\rho},\xi) + S_3(\mathbf{\rho},\xi) & S_1(\mathbf{\rho},\xi) + iS_2(\mathbf{\rho},\xi) \\ S_1(\mathbf{\rho},\xi) - iS_2(\mathbf{\rho},\xi) & S_0(\mathbf{\rho},\xi) - S_3(\mathbf{\rho},\xi) \end{pmatrix}, \qquad (53)$$

$$\hat{J} = \frac{1}{2}\begin{pmatrix} s_0 + s_3 & s_1 + is_2 \\ s_1 - is_2 & s_0 - s_3 \end{pmatrix}. \qquad (54)$$

Substituting (53) and (54) into (9), we obtain an expression for the CR Stokes vector:

$$\begin{pmatrix} S_0 \\ S_1 \\ S_2 \\ S_3 \end{pmatrix} = \begin{pmatrix} W_{00} + \frac{1}{2}(W_{11}+W_{-1-1}) & \mathrm{Re}[W_{01}+W_{0-1}] & \mathrm{Im}[W_{01}-W_{0-1}] & \frac{1}{2}(W_{11}-W_{-1-1}) \\ \mathrm{Re}[W_{01}+W_{0-1}] & W_{00}+\mathrm{Re}[W_{-11}] & \mathrm{Im}[W_{-11}] & \mathrm{Re}[W_{01}-W_{0-1}] \\ \mathrm{Im}[W_{01}-W_{0-1}] & \mathrm{Im}[W_{-11}] & W_{00}-\mathrm{Re}[W_{-11}] & \mathrm{Im}[W_{01}+W_{0-1}] \\ -\frac{1}{2}(W_{11}-W_{-1-1}) & -\mathrm{Re}[W_{01}-W_{0-1}] & -\mathrm{Im}[W_{01}+W_{0-1}] & W_{00}-\frac{1}{2}(W_{11}+W_{-1-1}) \end{pmatrix}_{\mathbf{\rho}_1=\mathbf{\rho}_2=\mathbf{\rho}} \begin{pmatrix} s_0 \\ s_1 \\ s_2 \\ s_3 \end{pmatrix}, \qquad (55)$$

where we use the property of the orbital correlation function, $W_{\nu\mu}(\mathbf{\rho},\mathbf{\rho},\xi)=W_{\mu\nu}(\mathbf{\rho},\mathbf{\rho},\xi)^*$, which is valid for $\mathbf{\rho}_1=\mathbf{\rho}_2=\mathbf{\rho}$. As a result, the CR intensity will be equal to:

$$I(\mathbf{\rho},\xi) = \left[W_{00} + \tfrac{1}{2}(W_{11}+W_{-1-1}),\ \mathrm{Re}[W_{01}+W_{0-1}],\ \mathrm{Im}[W_{0-1}-W_{01}],\ \tfrac{1}{2}(W_{-1-1}-W_{11})\right]_{\mathbf{\rho}_1=\mathbf{\rho}_2=\mathbf{\rho}} \cdot \mathbf{s}, \qquad (56)$$

where $\mathbf{s}=(s_0, s_1, s_2, s_3)$ is the Stokes vector of the input radiation. For unpolarized light, the input Stokes vector $\mathbf{s}=(1, 0, 0, 0)$, and Equation (56) becomes (10).

## 10. Appendix 2: Bessel-Gaussian Model of the CR

Following Reference [53], to calculate the integrals (34), it is necessary to transform the exponent $\exp[-i\rho_0 \kappa]$ under the integral sign. Let us transform it into the Bessel function of a complex argument using the asymptotic expansion of the Bessel function [82]:

$$i^\mu \exp[-i\rho_0\kappa] J_m(iq\kappa) \approx \frac{1}{\sqrt{2\pi i q\kappa}} \exp\left[-i(\rho_0+iq)\kappa - \frac{m^2}{2q\kappa} + i\frac{\pi}{2}\left(m+\mu+\frac{1}{2}\right)\right], \qquad (57)$$

and, since the Bessel function of the complex argument has the following asymptotic form:

$$J_{m+\mu}[(\rho_0+iq)\kappa] \approx \frac{1}{\sqrt{2\pi(\rho_0+iq)\kappa}} \exp\left[-i(\rho_0+iq)\kappa - i\frac{m^2}{2(\rho_0+iq)\kappa} + i\frac{\pi}{2}\left(m+\mu+\frac{1}{2}\right)\right], \qquad (58)$$

we can express (57) through (58) as:

$$i^\mu \exp[-i\rho_0\kappa] J_m(iq\kappa) \approx \sqrt{\frac{(\rho_0+iq)}{iq}} \exp\left(-\frac{m^2}{2q^2}\frac{\rho_0}{(\rho_0+iq)}\right) J_{m+\mu}[(\rho_0+iq)\kappa], \qquad (59)$$

where we use the property that the wave vectors near $\kappa=q$ make the main contribution to the integrand. As a result, substituting (59) into the expression for CR amplitude (34), the integrals over variable $\kappa$ are easily calculated, and we obtain:

$$C_\mu^{(+)}(\mathbf{q},\boldsymbol{\rho},\xi) = \frac{e^{i\mu\theta_q}}{2(1+i\xi)}\sqrt{\frac{-i\tau}{q}} \exp\left[-\frac{\rho^2+(\rho_0+q\xi)^2}{2(1+i\xi)} - iq\rho_0 - iq^2\frac{\xi}{2}\right] \times \\ \times \sum_{m=-\infty}^{+\infty} e^{i(m+\mu)(\varphi-\theta_q)-\frac{m^2}{2\varepsilon^2}\frac{1}{\rho_0\tau}} I_{m+\mu}\left(\frac{\rho(\rho_0+q\xi)}{1+i\xi}+iq\rho\right), \tag{60}$$

where $I_m(x)$ is the modified Bessel function of order $m$, $\tau=\rho_0+iq$ is a complex propagation parameter, and $\varepsilon=q/\rho_0$ is an asymmetry parameter (see the main text for details). To sum expression (60) over the OAM, we replace the exponential term with the Bessel functions of the complex argument:

$$\exp\left[-\frac{m^2}{2\varepsilon^2}\frac{1}{\rho_0\tau}\right] \approx \sqrt{2\pi\tau q\varepsilon}\exp\left[-\tau\rho_0\varepsilon^2\right] I_m\left(\tau\rho_0\varepsilon^2\right), \tag{61}$$

and we then use the Graf's addition theorem [73], obtaining the resulting formula (35).

## 11. Appendix 3: Coherence Degree of the CR Cones

To calculate the coherence degree of the CR cones in the limit of high spatial coherence, it is necessary to determine the explicit form of the amplitude and phase modulation (31). Since the random wave vector is small in this limit, we expand the amplitude and phase modulations in a Taylor series up to the second order in the random wave vector $\mathbf{q}$. Then, we substitute the resulting expansions into products of identical and different cones, obtaining the expressions:

$$\left|\mathbf{C}^{(\pm)}(\mathbf{q},\boldsymbol{\rho})\right|^2 = \left|\mathbf{C}^{(\pm)}(0,\boldsymbol{\rho})\right|^2 \exp\left[\pm 2\sum_i \frac{\partial A}{\partial q_i} q_i + \sum_{ij}\frac{\partial^2 A}{\partial q_i \partial q_j} q_i q_j\right], \tag{62}$$

$$\mathbf{C}^{(-)}(\mathbf{q},\boldsymbol{\rho})^*\mathbf{C}^{(+)}(\mathbf{q},\boldsymbol{\rho}) = \mathbf{C}^{(-)}(0,\boldsymbol{\rho})^*\mathbf{C}^{(+)}(0,\boldsymbol{\rho})\exp\left[\sum_{ij}\left(i\frac{\partial^2 B}{\partial q_i \partial q_j} + \frac{\partial^2 A}{\partial q_i \partial q_j}\right) q_i q_j\right]. \tag{63}$$

The intensity of the cones and the cross-cone correlation function from (62) and (63) are obtained by averaging over a random wave vector:

$$I^{(\pm)}(\boldsymbol{\rho}) = \left\langle\left|\mathbf{C}^{(\pm)}(\boldsymbol{\rho})\right|^2\right\rangle = \int\frac{d\mathbf{q}}{2\pi}\tilde{g}(\mathbf{q})\left|\mathbf{C}^{(\pm)}(\mathbf{q},\boldsymbol{\rho})\right|^2, \tag{64}$$

$$\left\langle\mathbf{C}^{(-)}(\boldsymbol{\rho})^*\mathbf{C}^{(+)}(\boldsymbol{\rho})\right\rangle = \int\frac{d\mathbf{q}}{2\pi}\tilde{g}(\mathbf{q})\,\mathbf{C}^{(-)}(\mathbf{q},\boldsymbol{\rho})^*\mathbf{C}^{(+)}(\mathbf{q},\boldsymbol{\rho}), \tag{65}$$

as well as using the property:

$$\left\langle\exp\left(\sum_i 2\alpha_i q_i + \sum_{ij}\beta_{ij} q_i q_j\right)\right\rangle = \exp\left(\sum_i (\alpha_i^2 + \beta_{ii}/2)\langle\mathbf{q}^2\rangle\right), \tag{66}$$

which is valid in the limit of small random wave vectors, as well as when the mean $<q_i>=0$ and variance $<q_i q_j>=<\mathbf{q}^2>\delta_{ij}/2$, where $\delta_{ij}$ is the Kronecker symbol. These conditions are satisfied if the weight function $\tilde{g}(\mathbf{q})$ does not depend on the angular variable $\theta_q$. As a result, we obtain the following:

$$I^{(\pm)}(\boldsymbol{\rho}) = \left|\mathbf{C}^{(\pm)}(0,\boldsymbol{\rho})\right|^2 \exp\left(\left[2(\nabla A)^2 + \Delta A\right]\langle\mathbf{q}^2\rangle/2\right), \tag{67}$$

$$\left\langle\mathbf{C}^{(-)}(\boldsymbol{\rho})^*\mathbf{C}^{(+)}(\boldsymbol{\rho})\right\rangle = \mathbf{C}^{(-)}(0,\boldsymbol{\rho})^*\mathbf{C}^{(+)}(0,\boldsymbol{\rho})\exp\left([i\Delta B + \Delta A]\langle\mathbf{q}^2\rangle/2\right), \tag{68}$$

where $\nabla$ and $\Delta$ are the Nabla and Laplace operators acting on the variable $\mathbf{q}$, respectively. From (67), it follows that the intensity of the cones will contain only contributions from the amplitude modulation $A$. Moreover, the contribution depending on $\Delta A$ enters into both expressions, while the phase modulation enters only into the cross-cone correlation function (68). We substitute (67) and (68) into the formula for coherence degree of the CR cones (39), obtaining the final formula (43):

$$\alpha_{CR} = \exp\left[\frac{1}{2}\frac{\partial^2 g}{\partial \rho^2}\rho_C^2\right] \approx 1 + \frac{1}{2}\frac{\partial^2 g}{\partial \rho^2}\rho_C^2 = g(\rho_C), \quad (69)$$

where we introduce the characteristic distance $\rho_C=2\nabla A$ and use the relationship between the average value of the square of the wave vector and the derivative of the input coherence degree $<q^2>=-2\partial^2 g(\rho)/\partial\rho^2$.

The next step is to calculate the characteristic distance $\rho_C$. To this end, we use formula (31), which includes the expression for the CR cone (29) in the focal plane ($\xi=0$):

$$C_\mu^{(+)}(\mathbf{q},\boldsymbol{\rho}) = \frac{1}{2}\int\frac{d\boldsymbol{\kappa}}{2\pi}\exp(i\boldsymbol{\kappa}\boldsymbol{\rho} - i\kappa\rho_0 + i\mu\theta_\kappa)\tilde{\mathcal{E}}^{(0)}(\boldsymbol{\kappa}-\mathbf{q}). \quad (70)$$

It greatly simplifies if we consider the limit of a well-defined CR ring $\rho_0 \gg 1$, as well as set the radial coordinate equal to the radius of the ring $\rho=\rho_0$. In this case, it is possible, using the stationary phase method, to integrate over the angle expression (70):

$$C_\mu^{(+)}(\mathbf{q},\rho_0,\varphi) = \frac{1}{\sqrt{2\pi\rho_0}}e^{i\mu\varphi - i\frac{\pi}{4}}\int_0^{+\infty}d\kappa\,\kappa^{1/2}\tilde{\mathcal{E}}^{(0)}(\sqrt{\kappa^2 - 2\kappa\mathbf{q}\mathbf{e}_r + q^2}), \quad (71)$$

where we use the property of the Fourier transform $\tilde{\mathcal{E}}^{(0)}(\kappa)$ to depend only on the radial component of the wave vector $\kappa$, which corresponds to the presence of radial symmetry in the field amplitude $\mathcal{E}^{(0)}(\rho)$. Therefore, because the modulus of the random wave vector is quite small ($q \ll \kappa$) in the limit of high spatial coherence, we use the expansion of the integrand (71) in a Taylor series:

$$C_\mu^{(+)}(\mathbf{q},\rho_0,\varphi) = \frac{1}{\sqrt{2\pi\rho_0}}e^{i\mu\varphi - i\frac{\pi}{4}}\int_0^{+\infty}d\kappa\left(\kappa^{1/2} + \kappa^{-1/2}\frac{\mathbf{q}\mathbf{e}_r}{2}\right)\tilde{\mathcal{E}}^{(0)}(\kappa), \quad (72)$$

where we also integrate by parts to express the answer in terms of the Fourier transform of the deterministic field amplitude $\tilde{\mathcal{E}}^{(0)}(\kappa)$. Knowing the expression for the CR cone (72), we can find the amplitude modulation $A$ using formula (31). Afterward, we easily obtain expression (42) for the characteristic distance $\rho_C=2\nabla A$, on which the input coherence degree in expression (69) depends.

Now, consider the case of low spatial coherence. The expression for the intensity of the cones (44) in this limit was obtained earlier in the text. Therefore, to find the coherence degree of the CR cones with formula (39), we only need to calculate the cross-cone correlation function. To this end, we use formula (35), assuming that $\rho \approx \rho_0 \gg 1$, and we can replace the modified Bessel function with its asymptotics, expressing the product of the different CR cones, in the following form:

$$\mathbf{C}^{(+)}(\mathbf{q},\boldsymbol{\rho})^*\mathbf{C}^{(-)}(\mathbf{q},\boldsymbol{\rho}) = \frac{-i\varepsilon\tau}{2\sqrt{|\boldsymbol{\rho}+\boldsymbol{\rho}_0\varepsilon^2||\boldsymbol{\rho}-\boldsymbol{\rho}_0\varepsilon^2|}}\exp\left[-\rho^2 - \rho_0^2 - 2iq\rho_0 - 2\tau\rho_0\varepsilon^2 + \tau|\boldsymbol{\rho}+\boldsymbol{\rho}_0\varepsilon^2| + \tau|\boldsymbol{\rho}-\boldsymbol{\rho}_0\varepsilon^2|\right], \quad (73)$$

where we assume that the radiation is unpolarized. From formula (73), it is clear that the product of the different cones has a maximum in the region $q \approx 1\text{-}2$, wherein the asymmetry parameter is $\varepsilon=q/\rho_0 \ll 1$. As a result, we can significantly simplify (73), obtaining:

$$\mathbf{C}^{(+)}(\mathbf{q},\rho_0,\varphi)^*\mathbf{C}^{(-)}(\mathbf{q},\rho_0,\varphi) = -i\frac{q}{2\rho_0}\exp\left[-2q^2\right]. \quad (74)$$

In addition, since the intensity of the cones (44) for $\rho=\rho_0$ will be equal to:

$$I^{(\pm)}(\rho_0,\varphi) = \frac{1}{2}\exp\left[-2\rho_0^2\right]I_0(2\rho_0^2) \approx 1/4\rho_0\sqrt{\pi}, \quad (75)$$

the degree of coherence (39) is transformed into the form:

$$\alpha_{CR} = 2\sqrt{\pi}\int_0^\infty dq\,\tilde{g}(q)\,q^2\exp\left[-2q^2\right]. \quad (76)$$

To express the coherence degree of the CR cones in terms of input coherence degree $g(\rho)$, we use the Hankel transform:

$$\tilde{g}(q) = \int_0^\infty d\rho \rho J_0(q\rho) g(\rho), \tag{77}$$

and we then substitute (77) into (76), thus obtaining (45).